\documentclass[a4paper, conference]{IEEEtran}

\usepackage{paralist}
\usepackage{eurosym}
\usepackage{flushend}
\usepackage{cite}
\usepackage{amsmath,graphicx}
\usepackage{amssymb}
\usepackage{url}
\usepackage[utf8]{inputenc}
\usepackage{amsfonts}
\usepackage[colorlinks=true]{hyperref}
\usepackage{soul}
\usepackage{caption}
\usepackage[list=true]{subcaption}
\usepackage{booktabs}
\usepackage{tabu}
\usepackage{wrapfig}
\usepackage{arydshln}
\usepackage[normalem]{ulem}
\usepackage{multirow}
\usepackage{siunitx}

\begin{document}

\title{Few-Shot Learning with Uncertainty-based Quadruplet Selection for Interference Classification in GNSS Data}

\author{\IEEEauthorblockN{Felix Ott\IEEEauthorrefmark{1}\IEEEauthorrefmark{3}\IEEEauthorrefmark{4},\thanks{\IEEEauthorrefmark{3}Felix Ott and Lucas Heublein contributed equally.}
    Lucas Heublein\IEEEauthorrefmark{1}\IEEEauthorrefmark{3},
    Nisha Lakshmana Raichur\IEEEauthorrefmark{1},
    Tobias Feigl\IEEEauthorrefmark{1}, 
    Jonathan Hansen\IEEEauthorrefmark{2}, \\
    Alexander Rügamer\IEEEauthorrefmark{2},
    Christopher Mutschler\IEEEauthorrefmark{1}}
  \IEEEauthorblockA{Fraunhofer Institute for Integrated Circuits IIS, Nuremberg, Germany \\
    \IEEEauthorrefmark{1}Precise Positioning and Analytics Department \& \IEEEauthorrefmark{2}Satellite-based Localization Systems Department}
  \IEEEauthorblockA{\IEEEauthorrefmark{4}~felix.ott@iis.fraunhofer.de}}

\IEEEoverridecommandlockouts
\IEEEpubid{\makebox[\columnwidth]{
979-8-3503-8078-1/24/\$31.00~\copyright2024
IEEE \hfill} \hspace{\columnsep}\makebox[\columnwidth]{ }}

\maketitle

\begin{abstract}
Jamming devices pose a significant threat by disrupting signals from the global navigation satellite system (GNSS), compromising the robustness of accurate positioning. Detecting anomalies in frequency snapshots is crucial to counteract these interferences effectively. The ability to adapt to diverse, unseen interference characteristics is essential for ensuring the reliability of GNSS in real-world applications. In this paper, we propose a few-shot learning (FSL) approach to adapt to new interference classes. Our method employs quadruplet selection for the model to learn representations using various positive and negative interference classes. Furthermore, our quadruplet variant selects pairs based on the aleatoric and epistemic uncertainty to differentiate between similar classes. We recorded a dataset at a motorway with eight interference classes on which our FSL method with quadruplet loss outperforms other FSL techniques in jammer classification accuracy with 97.66\%. \\
\href{https://gitlab.cc-asp.fraunhofer.de/darcy_gnss/FIOT_highway}{https://gitlab.cc-asp.fraunhofer.de/darcy\_gnss/FIOT\_highway}
\end{abstract}
\IEEEpeerreviewmaketitle

\section{Introduction}
\label{label_introduction}

Anomaly detection involves identifying deviations in data patterns from the anticipated norm \cite{nassif_talib}. Its applications span various domains such as fraud detection~\cite{dou_liu}, monitoring medical conditions~\cite{schlegl_seeboeck}, and GNSS-based applications~\cite{raichur_ion_gnss,merwe_franco,kim_mokaya}. The disruption of GNSS receivers' localization accuracy occurs due to interference signals from jammers. Notably, this issue has intensified in recent years with the proliferation of more affordable and accessible jamming devices \cite{merwe_franco_jdidi}. These jamming attacks result in critical consequences, such as collisions with self-driving cars~\cite{tesla_spoofing} or disruptions in airplane GPS systems~\cite{finland_reuters}. Hence, the detection and classification of potential interference signals are imperative \cite{brieger_ion_gnss}.

Both classical and machine learning (ML) methods have proven to detect and classify interference \cite{raichur_ion_gnss,merwe_franco,brieger_ion_gnss,yang_kang,swinney_woods}. However, the unpredictable emergence of novel, ``undetected'' jammer types necessitates rapid model adaptation. Few-shot learning (FSL) \cite{snell_swersky,wang_chao,vinyals_blundell,sung_yang_zhang,luo_wu_zhang,ye_hu_zhang,liu_song_qin,ziko_dolz} facilitates model adaptation to new classes by leveraging a minimal number of query samples. Numerous FSL methods exist that focus on the architecture \cite{vinyals_blundell,ye_hu_zhang}, label propagation \cite{liu_song_qin,ziko_dolz}, or the distance function \cite{sung_yang_zhang,snell_swersky}. Notably, within the domain of GNSS data, an issue arises concerning imbalanced data distributions, where interference-free signals markedly outnumber those containing interference, and certain interference classes are underrepresented \cite{ott_dissertation}. \textit{Contrastive learning} pairs positive and negative instances with corresponding anchors to enrich training \cite{chen_kornblith}. The \textit{triplet loss} jointly minimizes both pairs, thereby enhancing an improved feature representation \cite{schroff_kalenichenko}. In contrast, the \textit{quadruplet loss} achieves a more extensive inter-class variation and reduced intra-class variance \cite{chen_chen}. The quadruplet loss utilizes an additional label denoting a similar class. However, the selection of triplet and quadruplet pairs, particularly within unbalanced datasets, presents a challenge for optimizing the training process \cite{do_tran_reid}. While pair selection is addressed by broad research, a crucial gap remains unaddressed, i.e., quadruplet selection based on uncertainty quantification (UQ). Hence, our approach involves both aleatoric and epistemic UQ to select a similar sample corresponding to an anchor, positive, and negative pair.

The primary objective of this work is to adapt to new jammer classes observed within GNSS data. We propose the following contributions: (1) We present a snapshot-based GNSS dataset recorded at a motorway that contains various jammer types. (2) We benchmark FSL techniques and various adaptation layers for post-training for an optimal enhancement of additional classes. (3) We contribute a quadruplet loss function with uncertainty-based pair selection that outperforms the triplet loss. Therefore, we quantify both aleatoric and epistemic uncertainty to select challenging samples resembling similar class pairs. (4) Our evaluation on our dataset shows that our quadruplet loss leads to a continuous feature representation reducing false positive predictions, showcasing its efficacy in the context of GNSS interference analysis.

Section~\ref{label_related_work} summarizes related work and Section~\ref{label_background} gives a background on FSL and pairwise learning. In Section~\ref{label_method}, we propose our quadruplet selection method for FSL. We show details of our dataset in Section~\ref{label_dataset}. Section~\ref{label_evaluation} summarizes evaluation results and we conclude in Section~\ref{label_conclusion}.

\section{Related Work}
\label{label_related_work}

Yang et al.~\cite{yang_kang} used automatic gain control (AGC) and an adaptive notch filter for interference detection and characterization according to interference types and power levels. While Marcos et al.~\cite{marcos_caizzone} characterized GNSS interferences in maritime applications, Murrian et al.~\cite{murrian_narula} proposed a situational awareness monitoring for observation of terrestrial GNSS interference from low-earth orbit. For an overview of existing localization systems such as received signal strength (RSS), source angle of arrival (AoA), time difference of arrival (TDoA), and frequency difference of arrival (FDoA), refer to \cite{dempster_cetin}. Biswas et al.~\cite{biswas_cetin} proposed a particle filter for localizing a moving GNSS source using AoA and TDoA. Borio et al.~\cite{borio_closas} used an orthonormal transformation to project the received GNSS samples into an appropriate transform domain. van der Merwe et al.~\cite{merwe_franco,merwe_franco_jdidi} introduced a low-cost COTS GNSS interference monitoring, detection, and classification receiver. Swinney et al.~\cite{swinney_woods} considered the jamming signal power spectral density, spectrogram, raw constellation, and histogram signal representations as images to utilize transfer learning from the imagery domain. They evaluated convolutional neural networks (CNNs), i.e., VGG16, support vector machines (SVMs), logistic regression, and random forests for the classification task. Inspired by Swinney et al.~\cite{swinney_woods}, we also employ spectrograms. Ferre et al.~\cite{ferre_fuente} presented a SVM-based jammer classification method, however used a dataset with only five different jammer types. Also, \cite{li_huang_lang,xu_ying_li} used a twin SVM-based method. Mehr et al.~\cite{mehr_dovis} classified chirp signals utilizing a CNN with representation transformed with the Wigner-Ville and Fourier method.

Brieger et al.~\cite{brieger_ion_gnss} considered both the spatial and temporal relationships between samples when fusing ResNet18~\cite{he_zhang} (for data of a coarse-grained low-resolution low-cost sensor) and TS-Transformer (for data of a bandwidth-limited low-cost sensor) with a joint loss function and a late fusion technique. As ResNet18 proved to be robust for interference classification, we also utilize ResNet18 for feature extraction. Raichur et al.~\cite{raichur_ion_gnss} and Borio et al.~\cite{borio_gioia_stern} proposed a crowdsourcing approach with smartphone-based features to localize the source of any detected interference. Jdidi et al.~\cite{jdidi_brieger} proposed an unsupervised method that adapts to different environment-specific factors, i.e., multipath and dynamics, and variations in signal strength. In contrast to the prior work, we focus on improving the robustness of ML models when confronted with unbalanced datasets comprising eight interference classes, facilitating unsupervised adaptation to unknown classes. Therefore, we highlight on learning an improved representation discerning between (non-)interference classes. Subsequently, in Section~\ref{label_background_FSL}, we provide a background on FSL for class adaptation, followed by Section~\ref{label_background_PL}, which presents pairwise learning, specifically the quadruplet loss.

\section{Methodological Background}
\label{label_background}

\subsection{Background on Few-shot Learning (FSL)}
\label{label_background_FSL}

Within this section, we focus on the most relevant FSL methods. The interested reader finds a broader overview in \cite{song_wang_mondal} and a benchmark in \cite{luo_wu_zhang}. \textit{Matching nets} by Vinyals et al.~\cite{vinyals_blundell} emphasizes metric learning, i.e., leveraging the cosine distance \cite{ott_acmmm}, by mapping a small labelled support set and an unlabelled example to its label. \textit{LaplacianShot}~\cite{ziko_dolz} performs transductive Laplacian-regularized inference, encouraging nearby query samples to have consistent label assignments. Snell et al.~\cite{snell_swersky} proposed a \textit{prototypical network} (PN), where classification is executed by computing distances, i.e., the Euclidean distance, to prototype representations associated with each class. Liu et al.~\cite{liu_song_qin} presented a PN based on the cosine distance. The \textit{relation network}~\cite{sung_yang_zhang} computes relation scores between query images and a few examples of each new class without further network updates. We compare these methods using our GNSS-based dataset and select -- due to the highest F2-score -- the PN as the baseline for FSL in conjunction with our uncertainty-based \cite{klass_lorenz_strl} contrastive learning method.

\subsection{Background on Pairwise Learning}
\label{label_background_PL}

The pairwise contrastive loss minimizes the distance between pairs of feature embeddings belonging to the same class and maximizes the distance between pairs of feature embeddings from different classes. The contrastive loss benefits from larger batch sizes and more training iterations compared to supervised learning \cite{chen_kornblith}. A limitation is that the optimization of positive pairs occurs independently of negative pairs \cite{do_tran_reid}. The triplet loss \cite{schroff_kalenichenko} addresses this issue by defining an anchor, a positive, and a negative point. It aims to enforce the positive pair distance to be smaller than the negative pair distance by a certain margin \cite{ott_access}. The quadruplet loss requires an additional \textit{semi-hard} negative sample, which is closer to the anchor than the original negative sample but remains dissimilar. This results in a model output with a greater inter-class variation and a smaller intra-class variation compared to the triplet loss. Despite contrastive learning not being applied to GNSS-based interference detection and classification, we utilize the quadruplet loss for an improved, continuous feature representation of jammer classes.

\begin{figure*}[!t]
    \centering
    \includegraphics[width=1.0\linewidth]{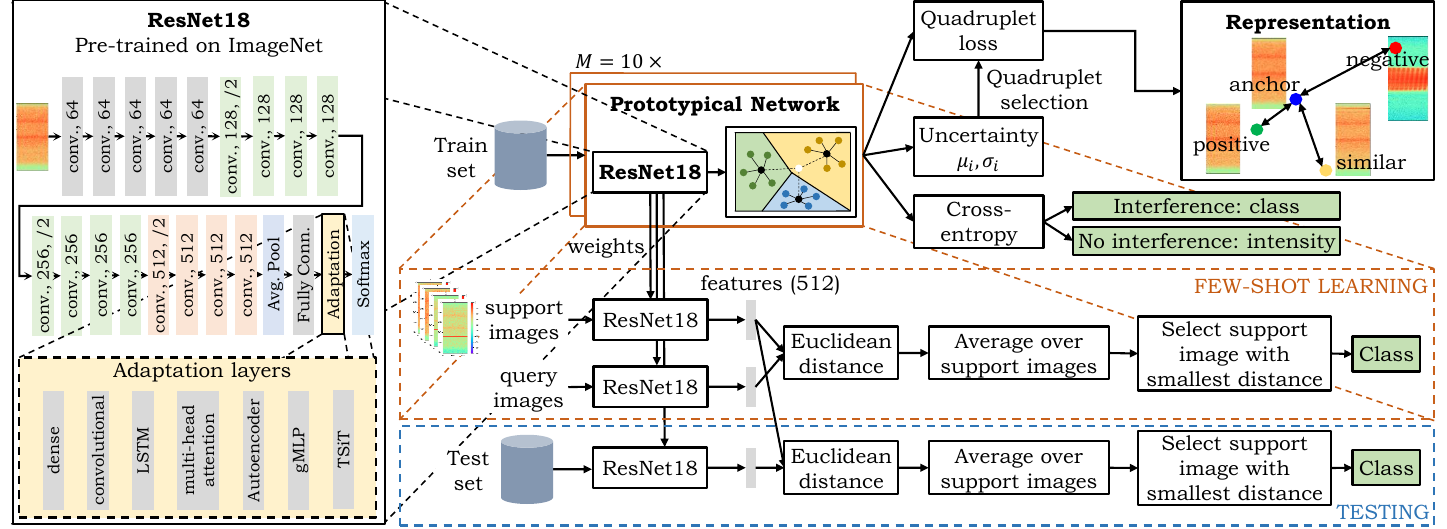}
    \caption{Overview of our FSL method with uncertainty-based quadruplet selection to adapt to unseen interference classes.}
    \label{figure_method_overview}
\end{figure*}

However, this raises the computational challenge of runtime complexity for large training sets \cite{do_tran_reid}. Hence, several approaches exist to mitigate this complexity through optimal triplet/quadruplet mining. Given that hard negative samples often share relatively similar characteristics with anchors and positive samples within a batch, Lv et al.~\cite{lv_gu_liu} introduced an element-weighted loss to expand the distance between partial elements of feature vectors. We face a similar challenge in the unbalanced GNSS-based dataset due to the similarity in background intensity between positive and negative samples. We mitigate this problem by utilizing the model's uncertainty to select suitable quadruplets, thereby enhancing the quality of the feature representation. However, UQ for sample selection remains unexplored. Warburg et al.~\cite{warburg_jorgensen} proposed a likelihood-based method that aligns with the triplet constraint, evaluating the probability of an anchor being closer to a positive sample than a negative one. For domain adaptation, Zheng et al.~\cite{zheng_lan_zeng} utilized the credibility of assigned pseudo-labels to mitigate the impact of noisy labels, thereby adjusting their influence on the contrastive loss function. Wu et al.~\cite{wu_yang_gu} incorporated the triplet loss for medical image analysis and quantified predictive uncertainty; however, selected semi-hard samples without factoring in predicted uncertainty.

\section{Methodology}
\label{label_method}

First, we provide an overview of our FSL method in Section~\ref{label_rw_overview}. Next, we propose our quadruplet loss functions in Section~\ref{label_rw_cl}. In Section~\ref{label_rw_uq}, we present details about quantifying uncertainty for quadruplet selection.

\subsection{Method Overview}
\label{label_rw_overview}

First, we define notations. Let $\mathbf{X} \in \mathbb{R}^{h \times w}$ with entries $x_{i,j} \in [0, 255]$ and height $h$ and width $w$ represent an image from the image training set. The images are obtained by calculating the magnitude spectrogram of a snapshot of IQ-samples. The image training set is a subset of the array $\mathcal{X} = \{\mathbf{X}_1,\ldots,\mathbf{X}_{n_X}\} \in \mathbb{R}^{n_X \times h \times w}$, where $n_X$ is the number of images in the training set. The goal is to predict an unknown class label $y \in \Omega$. We define the classification task as a binary task with the classes '\textit{no interference}' (0) and '\textit{interference}' (1), or as a multi-class task with three '\textit{no interference}' classes (0, 1, and 2) with varying background intensity and eight '\textit{interference}' classes (3 to 10), e.g., '\textit{chirp}' \cite{brieger_ion_gnss}. The goal is to learn representative feature embeddings $f(\mathbf{X}) \in \mathbb{R}^{q \times t}$ to map the image input into a feature space $\mathbb{R}^{q \times t}$, where $f$ is the output of a specified model layer and $q \times t$ is the dimension of the layer output \cite{ott_access}.

We give an overview of our FSL method in Figure~\ref{figure_method_overview}. First, we train a FSL model, here the PN~\cite{snell_swersky}, employing cross-entropy as the baseline criterion, to learn a representation with a small intra-class distance and high inter-class distance. The PN is based on ResNet18~\cite{he_zhang}, pre-trained using ImageNet. During the adaptation phase, we compute the Euclidean distance between the features of size 512 from the output of the final layer for both the support and query images. At inference, we select the class assigned to the query image, determined by the smallest distance computed to the average of the support images. As the main focus of this paper is to enhance the learned feature representation, we utilize the contrastive, triplet, and quadruplet loss functions, enabling the use of negative, positive, and similar samples (see top right of Figure~\ref{figure_method_overview}). For more details, refer to Section~\ref{label_rw_cl}. To select challenging quadruplets characterized by low confidence, we compute both aleatoric and epistemic uncertainties for each sample (refer to Section~\ref{label_rw_uq}). For evaluation, we compare our quadruplet loss and FSL techniques~\cite{snell_swersky,wang_chao,vinyals_blundell,sung_yang_zhang,luo_wu_zhang,ye_hu_zhang,liu_song_qin,ziko_dolz}, as well as ResNet18 augmented with various adaptation layers for post-training, i.e., dense layers, convolutional layers, long short-term memory (LSTM), multi-head attention, Autoencoder, gMLP~\cite{liu_dai_so_le}, and the transformer TSiT~\cite{zerveas_jayaraman_patel}.

\subsection{Pairwise Learning}
\label{label_rw_cl}

\begin{figure*}[!t]
    \centering
	\begin{minipage}[t]{0.085\linewidth}
        \centering
    	\includegraphics[trim=216 34 60 38, clip, width=1.0\linewidth]{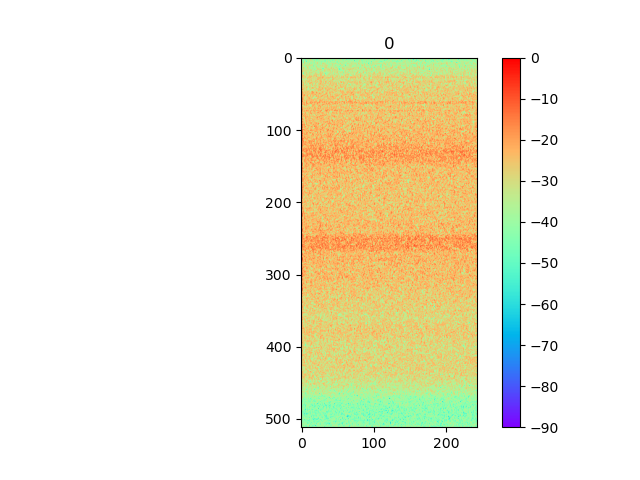}
    	\subcaption{Class 0.}
    	\label{figure_exemplary_samples0}
    \end{minipage}
    \hfill
	\begin{minipage}[t]{0.085\linewidth}
        \centering
    	\includegraphics[trim=216 34 60 38, clip, width=1.0\linewidth]{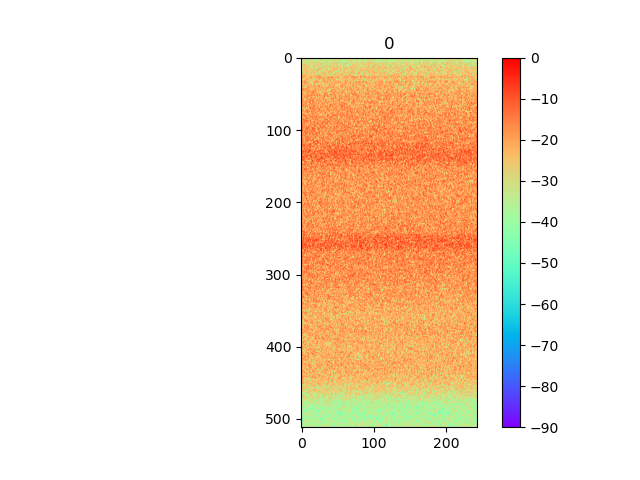}
    	\subcaption{Class 1.}
    	\label{figure_exemplary_samples1}
    \end{minipage}
    \hfill
	\begin{minipage}[t]{0.085\linewidth}
        \centering
    	\includegraphics[trim=216 34 60 38, clip, width=1.0\linewidth]{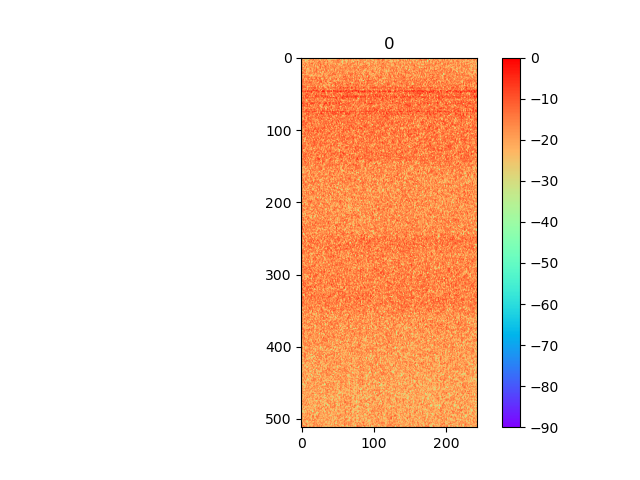}
    	\subcaption{Class 2.}
    	\label{figure_exemplary_samples2}
    \end{minipage}
    \hfill
	\begin{minipage}[t]{0.085\linewidth}
        \centering
    	\includegraphics[trim=216 34 60 38, clip, width=1.0\linewidth]{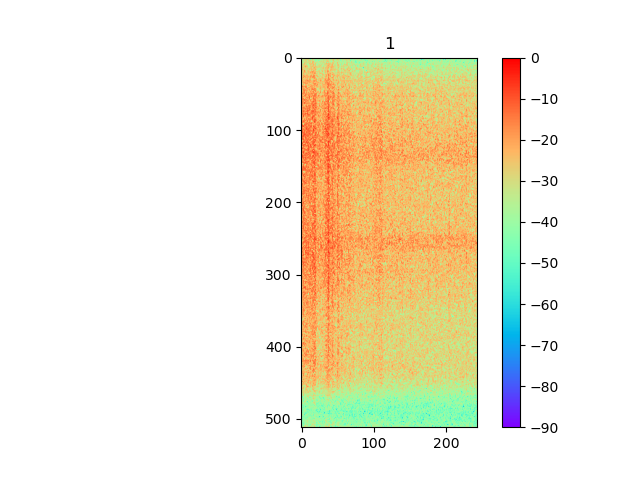}
    	\subcaption{Class 3.}
    	\label{figure_exemplary_samples3}
    \end{minipage}
    \hfill
	\begin{minipage}[t]{0.085\linewidth}
        \centering
    	\includegraphics[trim=216 34 60 38, clip, width=1.0\linewidth]{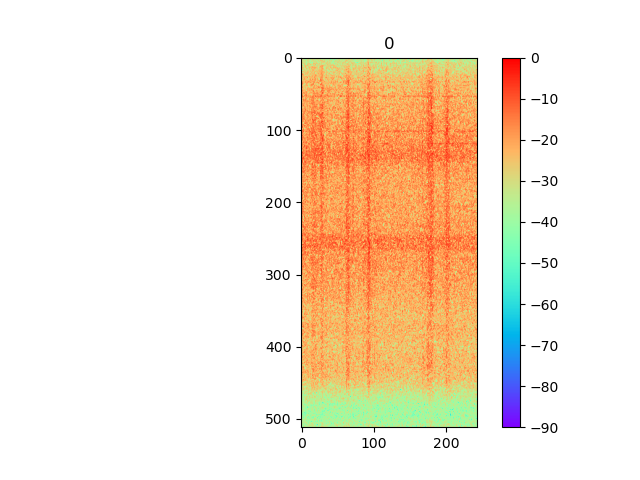}
    	\subcaption{Class 4.}
    	\label{figure_exemplary_samples4}
    \end{minipage}
    \hfill
	\begin{minipage}[t]{0.085\linewidth}
        \centering
    	\includegraphics[trim=216 34 60 38, clip, width=1.0\linewidth]{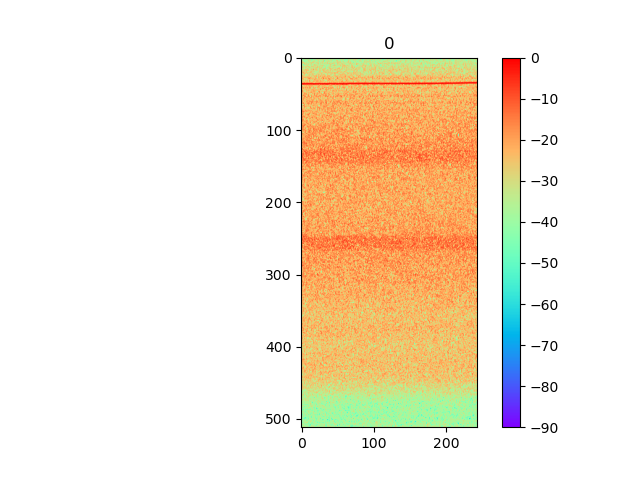}
    	\subcaption{Class 5.}
    	\label{figure_exemplary_samples5}
    \end{minipage}
    \hfill
	\begin{minipage}[t]{0.085\linewidth}
        \centering
    	\includegraphics[trim=216 34 60 38, clip, width=1.0\linewidth]{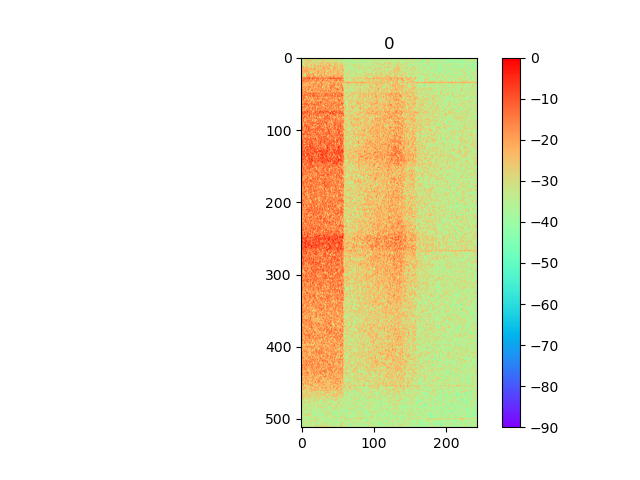}
    	\subcaption{Class 6.}
    	\label{figure_exemplary_samples6}
    \end{minipage}
    \hfill
	\begin{minipage}[t]{0.085\linewidth}
        \centering
    	\includegraphics[trim=216 34 60 38, clip, width=1.0\linewidth]{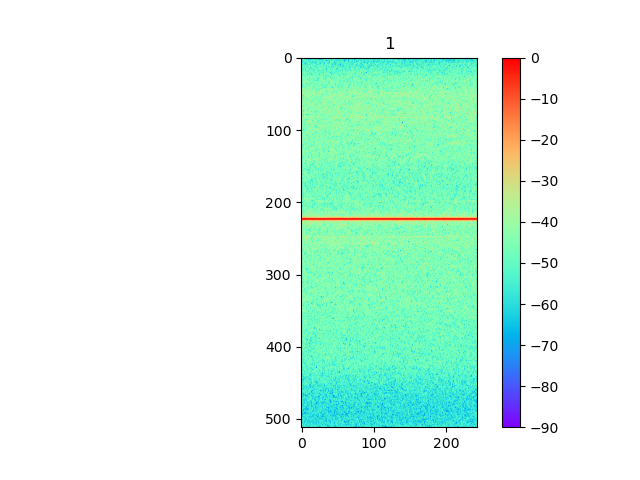}
    	\subcaption{Class 7.}
    	\label{figure_exemplary_samples7}
    \end{minipage}
    \hfill
	\begin{minipage}[t]{0.085\linewidth}
        \centering
    	\includegraphics[trim=216 34 60 38, clip, width=1.0\linewidth]{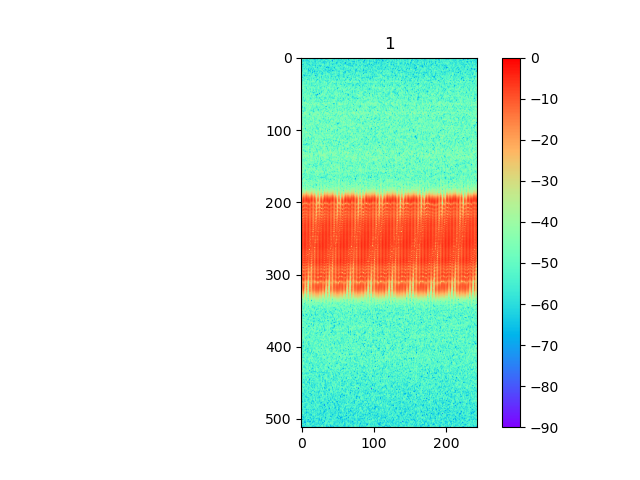}
    	\subcaption{Class 8.}
    	\label{figure_exemplary_samples8}
    \end{minipage}
    \hfill
	\begin{minipage}[t]{0.085\linewidth}
        \centering
    	\includegraphics[trim=216 34 60 38, clip, width=1.0\linewidth]{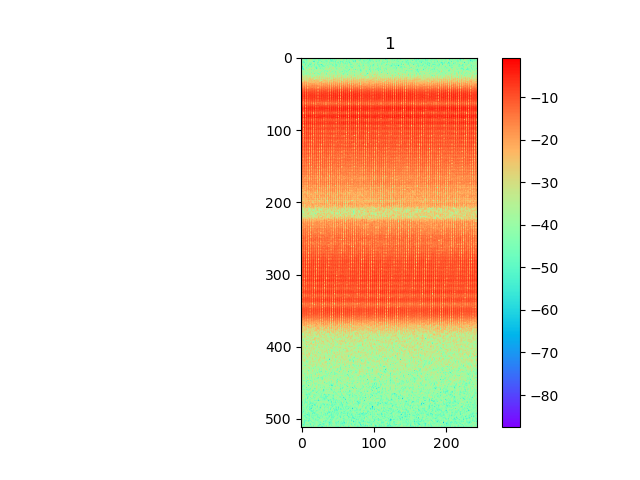}
    	\subcaption{Class 9.}
    	\label{figure_exemplary_samples9}
    \end{minipage}
    \hfill
	\begin{minipage}[t]{0.088\linewidth}
        \centering
    	\includegraphics[trim=216 34 60 38, clip, width=0.94\linewidth]{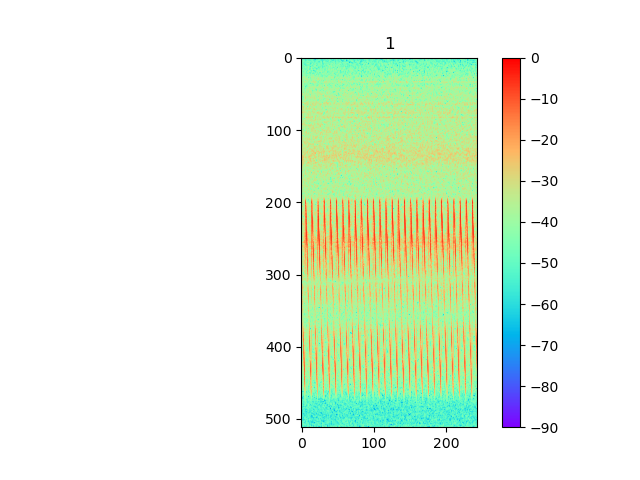}
    	\subcaption{Class 10.}
    	\label{figure_exemplary_samples10}
    \end{minipage}
    \caption{Exemplary spectrogram samples without interference (classes 0 to 2) and with interference (classes 3 to 10) between intensity [0, -90] with logarithmic scale. The x-axis shows the time in $\text{ms}$. The y-axis shows the frequency in $\text{MHz}$.}
    \label{figure_exemplary_samples}
\end{figure*}

\begin{figure}[!t]
    \centering
	\begin{minipage}[t]{0.328\linewidth}
        \centering
    	\includegraphics[width=1.0\linewidth]{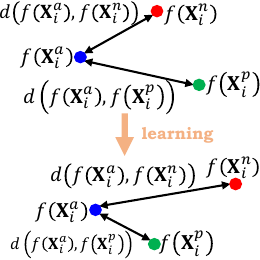}
    	\subcaption{Contrastive loss.}
    	\label{figure_contrastive_loss}
    \end{minipage}
    \hfill
	\begin{minipage}[t]{0.32\linewidth}
        \centering
    	\includegraphics[width=1.0\linewidth]{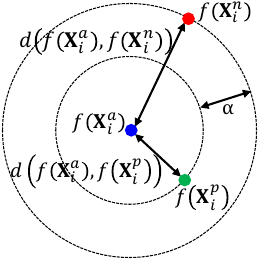}
    	\subcaption{Triplet loss.}
    	\label{figure_triplet_loss}
    \end{minipage}
    \hfill
	\begin{minipage}[t]{0.32\linewidth}
        \centering
    	\includegraphics[width=1.0\linewidth]{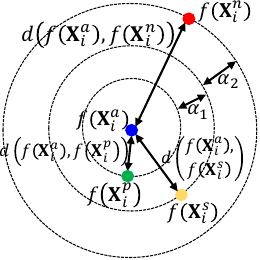}
    	\subcaption{Quadruplet loss.}
    	\label{figure_quadruplet_loss}
    \end{minipage}
    \caption{Overview of the pairwise learning loss functions.}
    \label{figure_loss_functions}
\end{figure}

Pairwise learning involves pairs with distinct labels and contributes to enhancing the training process by enforcing a margin between pairs of images with different identities. Consequently, the feature embedding for a specific label lives on a manifold while still ensuring sufficient discriminability, i.e., distance, from other identities \cite{ott_access,do_tran_reid}. Thus, we define an \textit{anchor} sample $\mathbf{X}_i^a$ from a specific label, a positive sample $\mathbf{X}_i^p$ from the same label, and a negative sample $\mathbf{X}_i^n$ from a different label. For all training samples $\big(f(\mathbf{X}_i^a), f(\mathbf{X}_i^p), f(\mathbf{X}_i^n) \big) \in \Phi$, our objective is to satisfy the following inequality:
\begin{equation}
\label{equ_contrastive}
    d \big( f(\mathbf{X}_i^a), f(\mathbf{X}_i^p) \big) + \alpha < d \big(f(\mathbf{X}_i^a), f(\mathbf{X}_i^n) \big),
\end{equation}
where $d$ is a distance function, here the Euclidean distance, $\alpha$ is a margin between positive and negative pairs, and $\Phi$ is the set of all possible pairs in the training set with $N$ being the number of all pairs. A softmax layer is adopted to normalize the feature embeddings $f$. The \textit{contrastive} loss \cite{chen_kornblith} minimizes the distance between the anchor and the positive sample, while separately maximizes the distance to the negative sample \cite{ott_access} (refer to Figure~\ref{figure_contrastive_loss}). The \textit{triplet} loss \cite{schroff_kalenichenko} (see Figure~\ref{figure_triplet_loss}) jointly optimizes the distances and its formulation can be expressed as follows:
\begin{equation}
\label{equ_triplet}
\begin{aligned}
    \mathcal{L}_{\text{triplet}}(\mathbf{X}^a, \mathbf{X}^p, \mathbf{X}^n) = \sum_{i=1}^N &\max \Big[ d \big(f(\mathbf{X}_i^a), f(\mathbf{X}_i^p)\big) - \\ &d \big(f(\mathbf{X}_i^a), f(\mathbf{X}_i^n)\big) + \alpha, 0 \Big].
\end{aligned}
\end{equation}
The quadruplet loss \cite{chen_chen} introduces a new constraint and is defined for all quadruplets $\big(f(\mathbf{X}_i^a), f(\mathbf{X}_i^{p}), f(\mathbf{X}_i^{s}), f(\mathbf{X}_i^n) \big) \in \Phi$ as 
\begin{equation}
\label{equ_qudruplet}
\begin{aligned}
    &\mathcal{L}_{\text{quadruplet}}(\mathbf{X}^a, \mathbf{X}^{p}, \mathbf{X}^{s}, \mathbf{X}^n) = \\ \sum_{i=1}^N &\max \Big[ d \big(f(\mathbf{X}_i^a), f(\mathbf{X}_i^{p})\big) - d \big(f(\mathbf{X}_i^a), f(\mathbf{X}_i^s)\big) + \alpha_1, 0 \Big] + \\ \sum_{i=1}^N &\max \Big[ d \big(f(\mathbf{X}_i^a), f(\mathbf{X}_i^{s})\big) - d \big(f(\mathbf{X}_i^a), f(\mathbf{X}_i^n)\big) + \alpha_2, 0 \Big],
\end{aligned}
\end{equation}
where $\alpha_1$ and $\alpha_2$ are the margins, and $\mathbf{X}^{s}$ is a similar sample to $\mathbf{X}^{a}$ or $\mathbf{X}^n$ (see Figure~\ref{figure_quadruplet_loss}). By enforcing this constraint, the minimum inter-class distance is required to exceed the maximum intra-class distance, irrespective of whether pairs contain the same label \cite{chen_chen}. However, it is challenging to pre-define a suitable margin threshold. A small threshold would yield few hard samples. As solely hard samples are utilized to train the model, few hard samples would lead to slow convergence and potentially result in the model with a suboptimal solution. Instead, a large threshold would generate an excessive number of hard training samples, potentially causing overfitting \cite{chen_chen}. We perform a hyperparameter search to determine optimal margins. We selectively choose similar samples from which the training process can derive a benefit.

\subsection{Uncertainty Quantification}
\label{label_rw_uq}

To curate pairs of anchor, positive, similar, and negative samples, we quantify the aleatoric and epistemic uncertainties for each sample described in the following. The models exhibit higher uncertainty for samples that bear resemblance to the positive sample, leading to a decrease in classification accuracy. The objective is to address this scenario by leveraging an enhanced representation facilitated by the quadruplet loss. Therefore, we apply Bayesian inference. The objective is to compute the posterior distribution $p(\theta|D)$, representing the neural network (NN) weights, given the training dataset $D$ and the model parameters $\theta$. As the computation of the posterior is usually intractable, a (local) approximation is often used \cite{klass_lorenz_strl}. We achieve this utilizing Deep Ensembles \cite{lakshminarayanan}, which is a committee of $M$ individual NNs initialized with a different seed. The initialization stands as the sole source of stochasticity in the model parameters, which are otherwise deterministic. The results are obtained by aggregating the predictions derived from $M=10$ independently trained NNs.

Next, we decompose the uncertainty: (1) \textit{Aleatoric} uncertainty represents stochasticity inherent in the data. (2) \textit{Epistemic} uncertainty is the model uncertainty, which can be reduced to zero for an increasing amount of observations. Our goal is to select samples with high epistemic uncertainty to form the quadruplet pairs, thereby reducing the model's uncertainty within the context of FSL. We adopt the decomposition approach by Kwon et al.~\cite{kwon_won} based on the variability of the softmax output. The predictive uncertainty can be specified as
\begin{equation}
\label{equ_kwon}
    \underbrace{\frac{1}{T} \sum_{t=1}^{T} \text{diag}(\hat{c}_t) - \hat{c}_t\hat{c}_t^\top}_{\text{aleatoric uncertainty}} + \underbrace{\frac{1}{T} \sum_{t=1}^{T} (\hat{c}_t - \bar{c})(\hat{c}_t - \bar{c})^\top}_{\text{epistemic uncertainty}},
\end{equation}
with $\hat{c}_t = (\hat{c}_{t,1}, \ldots ,\hat{c}_{t,K}) \in [0,1]^K$ being the softmax output of the NN based on one forward pass (out of $T$ stochastic forward passes), the number of classes $K$, $\sum_{i=1}^{K} \hat{c}_{t,i} = 1$, and $\bar{c} = \frac{1}{T} \sum_{t=1}^{T} \hat{c}_t$ \cite{klass_lorenz_strl,kwon_won}.

\section{Dataset}
\label{label_dataset}

We developed a hardware setup that captures short, wideband snapshots in both E1 and E6 GNSS bands. This setup is mounted to a bridge over a motorway. The setup records 20\,ms raw IQ snapshots triggered from the energy with a sample rate of 62.5\,MHz, an analog bandwidth of 50\,MHz and an 8\,bit bit-width. Further technical details can be found in \cite{brieger_ion_gnss}. Figure~\ref{figure_exemplary_samples} shows exemplary snapshots of the spectrogram. At certain frequencies the GPS/Galileo or GLONASS signals can easily be seen as a slight increase in the spectrum. Note that experts manually analyzed the datastreams by thresholding CN/0 and AGC values. Manual labeling of these snapshots has resulted in 11 classes: Classes 0 to 2 represent samples with no interferences, distinguished by variations in background intensity, while classes 3 to 10 contain different interferences. For instance, Figure~\ref{figure_exemplary_samples10} showcases a snapshot containing a potential chirp jammer type. Table~\ref{table_number_samples} presents a summary of the sample distributions across all 11 classes. This shows the dataset's imbalance, emphasizing the under-representation of positive class labels. The challenge lies in adapting to positive class labels with only a limited number of samples available. We partition the dataset into a 64\% training set, 16\% validation set, and a 20\% test set split (balanced over the classes). Utilizing the predicted uncertainty, outlined in Section~\ref{label_rw_uq}, we identify the highest levels of epistemic uncertainty among the subsequent class combinations that serve as our basis for quadruplet selection: $0 \rightarrow \{1, 3, 5, 7\}$, $1 \rightarrow \{0, 2, 4, 5\}$, $2 \rightarrow \{1\}$, $3 \rightarrow \{0, 4, 6\}$, $4 \rightarrow \{1, 3\}$, $5 \rightarrow \{0, 1\}$, $6 \rightarrow \{3\}$, $7 \rightarrow \{0, 8\}$, $8 \rightarrow \{7, 10\}$, $9 \rightarrow \{10\}$, and $10 \rightarrow \{8, 9\}$.

\begin{table}[t!]
\begin{center}
\setlength{\tabcolsep}{3.3pt}
    \caption{Overview of classes and sample numbers.}
    \label{table_number_samples}
    \vspace{-0.1cm}
    \small \begin{tabular}{ p{1.3cm} | p{0.5cm} | p{0.5cm} }
    \multicolumn{1}{c|}{\textbf{Class}} & \multicolumn{1}{c|}{\textbf{Label}} & \multicolumn{1}{c}{\textbf{Sample number}} \\ \hline
    \multicolumn{1}{l|}{Negative (\textit{low intensity})} & \multicolumn{1}{r|}{0} & \multicolumn{1}{r}{9,980} \\
    \multicolumn{1}{l|}{Negative (\textit{medium intensity})} & \multicolumn{1}{r|}{1} & \multicolumn{1}{r}{132,974} \\
    \multicolumn{1}{l|}{Negative (\textit{high intensity})} & \multicolumn{1}{r|}{2} & \multicolumn{1}{r}{54,620} \\
    \multicolumn{1}{l|}{Positive (\textit{pulsed})} & \multicolumn{1}{r|}{3} & \multicolumn{1}{r}{13} \\
    \multicolumn{1}{l|}{Positive (\textit{pulsed})} & \multicolumn{1}{r|}{4} & \multicolumn{1}{r}{28} \\
    \multicolumn{1}{l|}{Positive (\textit{out-of-band tone})} & \multicolumn{1}{r|}{5} & \multicolumn{1}{r}{59} \\
    \multicolumn{1}{l|}{Positive (\textit{noise})} & \multicolumn{1}{r|}{6} & \multicolumn{1}{r}{9} \\
    \multicolumn{1}{l|}{Positive (\textit{tone})} & \multicolumn{1}{r|}{7} & \multicolumn{1}{r}{39} \\
    \multicolumn{1}{l|}{Positive (\textit{chirp})} & \multicolumn{1}{r|}{8} & \multicolumn{1}{r}{79} \\
    \multicolumn{1}{l|}{Positive (\textit{two chirps})} & \multicolumn{1}{r|}{9} & \multicolumn{1}{r}{10} \\
    \multicolumn{1}{l|}{Positive (\textit{chirp})} & \multicolumn{1}{r|}{10} & \multicolumn{1}{r}{16} \\
    \end{tabular}
    \vspace{-0.2cm}
\end{center}
\end{table}

\section{Evaluation}
\label{label_evaluation}

For all experiments, we use Nvidia Tesla V100-SXM2 GPUs with 32 GB VRAM equipped with Core Xeon CPUs and 192 GB RAM. We use the vanilla SGD optimizer with a learning rate of $0.01$, decay of $0.0005$, and train the standard models for 100 epochs and the models with pairwise learning for 200 epochs. First, we present baseline results of FSL and for different layer adaptations. Next, we propose an evaluation of the uncertainty computation and our triplet and quadruplet loss with hyperparameter searches. In general, we train all possible class permutations for pre-training, adaptation, and evaluation, but present only results for the specific permutation (train classes: 0, 1, 2, 4, 5, 8, and 11; adaptation classes: 3, 7, 9, and 10). However, our statements can be generalized to all permutations. We train each method $M=10$ times and present the mean and standard variance F2-score.

\begin{figure}[!t]
    \centering
    \includegraphics[trim=11 11 10 11, clip, width=0.85\linewidth]{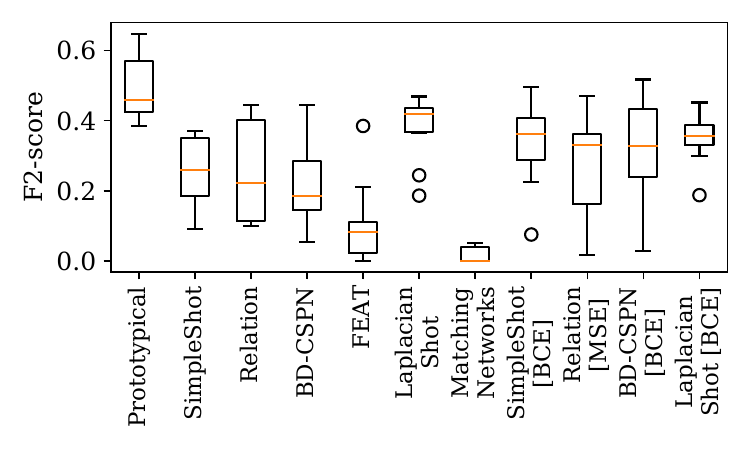}
    \vspace{-0.1cm}
    \caption{Baseline results (F2-score) of FSL methods.}
    \label{figure_baseline_fsl}
\end{figure}

\textbf{FSL Results.} The results of various FSL methods~\cite{snell_swersky,wang_chao,vinyals_blundell,sung_yang_zhang,luo_wu_zhang,ye_hu_zhang,liu_song_qin,ziko_dolz} are illustrated in Figure~\ref{figure_baseline_fsl}. Among these, prototypical network (PN)~\cite{snell_swersky} with Euclidean distance achieves the highest F2-scores. LaplacianShot~\cite{ziko_dolz}, SimpleShot~\cite{wang_chao} with binary CE (BCE) and cosine distance, and relation network~\cite{sung_yang_zhang} achieve high results ($\sim 0.4$ F2-score). However, a noticeable decline in F2-score is evident for the remaining methods. Generally, the CE loss proves to be more efficacious than the BCE. Consequently, our proposed method and subsequent discussions are grounded in the utilization of the PN with Euclidean distance.

\begin{figure}[!t]
    \centering
    \vspace{-0.2cm}
    \includegraphics[trim=10 91 71 124, clip, width=1.0\linewidth]{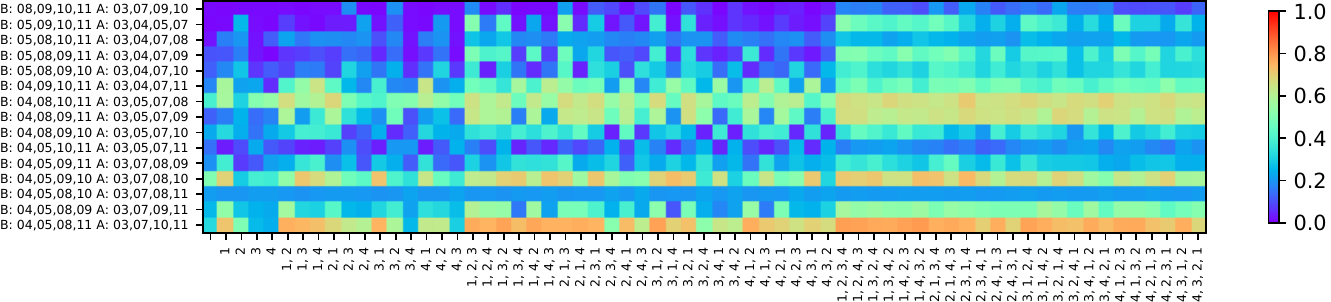}
    \caption{Evaluation results (F2-score) of all adaptation class permutations of the post-training with the \textit{convolutional} adaptation layer. y-axis: baseline (B) and adaptation (A) classes. x-axis: index of the additional classes.}
    \label{figure_post_train_dense}
    \vspace{-0.2cm}
\end{figure}

\textbf{Layer Adaptation.} Subsequently, we assess the adaptation layers that we post-train with the ResNet18 architecture, as depicted in Figure~\ref{figure_method_overview}. As an overview, Figure~\ref{figure_post_train_dense} provides a comprehensive summary of results across all class permutations, particularly highlighting the convolutional layer, which exhibits the most robust generalization. Notably, there is an increase in F2-score to all conceivable classes. The dense and LSTM layers also demonstrate a robust generalization. In contrast, multi-head attention, gMLP, and TSiT exhibit effectiveness primarily for specific permutations. The Autoencoder, while resembling the convolutional layer, manifests a lower F2-score. In summary, the findings indicate that simpler layers outperform more intricate models. Notably, ResNet18 inherently extracts robust features, rendering post-training only partially imperative.

\begin{figure}[!t]
    \centering
	\begin{minipage}[t]{1.0\linewidth}
	\begin{minipage}[t]{0.493\linewidth}
        \centering
    	\includegraphics[trim=97 46 703 83, clip, width=1.0\linewidth]{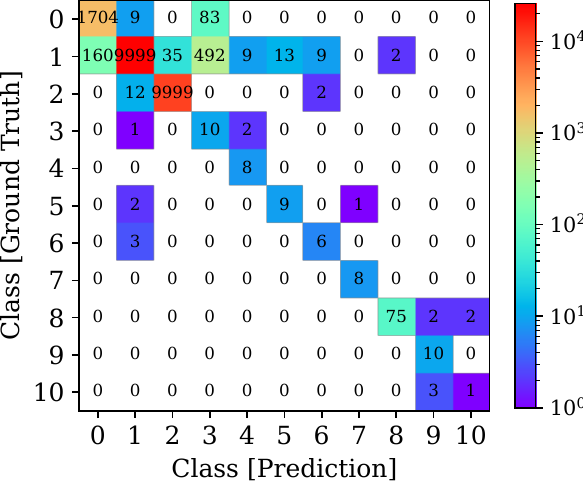}
    \end{minipage}
    \hfill
	\begin{minipage}[t]{0.493\linewidth}
        \centering
    	\includegraphics[trim=97 46 703 83, clip, width=1.0\linewidth]{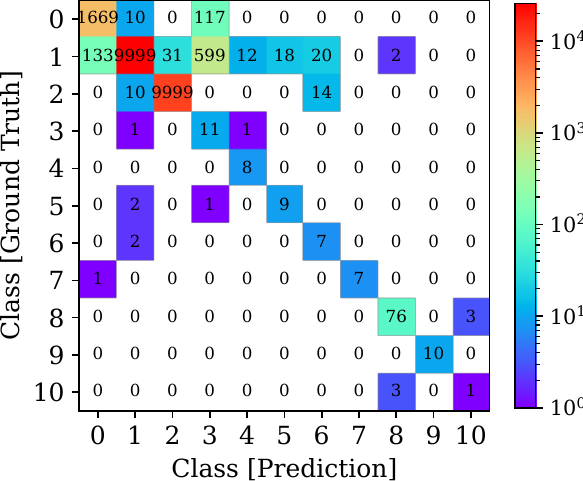}
    \end{minipage}
    \vspace{-0.6cm}
    \subcaption{Confusion matrix.}
    \label{figure_confusion_matrices1}
    \end{minipage}
	\begin{minipage}[t]{1.0\linewidth}
	\begin{minipage}[t]{0.493\linewidth}
        \centering
    	\includegraphics[trim=392 46 402 81, clip, width=1.0\linewidth]{images/000000_plot_015.pdf}
    \end{minipage}
    \hfill
	\begin{minipage}[t]{0.493\linewidth}
        \centering
    	\includegraphics[trim=392 46 402 81, clip, width=1.0\linewidth]{images/2.pdf}
    \end{minipage}
    \vspace{-0.6cm}
    \subcaption{Aleatoric uncertainty.}
    \label{figure_confusion_matrices2}
    \end{minipage}
	\begin{minipage}[t]{1.0\linewidth}
	\begin{minipage}[t]{0.493\linewidth}
        \centering
    	\includegraphics[trim=687 46 100 81, clip, width=1.0\linewidth]{images/000000_plot_015.pdf}
    \end{minipage}
    \hfill
	\begin{minipage}[t]{0.493\linewidth}
        \centering
    	\includegraphics[trim=687 46 100 81, clip, width=1.0\linewidth]{images/2.pdf}
    \end{minipage}
    \vspace{-0.6cm}
    \subcaption{Epistemic uncertainty.}
    \label{figure_confusion_matrices3}
    \end{minipage}
    \vspace{-0.1cm}
    \caption{Confusion matrices and uncertainties for the PN (left) and the PN with quadruplet loss (right) averaged over 10 runs.}
    \label{figure_confusion_matrices}
    \vspace{-0.3cm}
\end{figure}

\begin{figure}[!t]
    \centering
	\begin{minipage}[t]{0.493\linewidth}
        \centering
    	\includegraphics[trim=12 11 12 11, clip, width=1.0\linewidth]{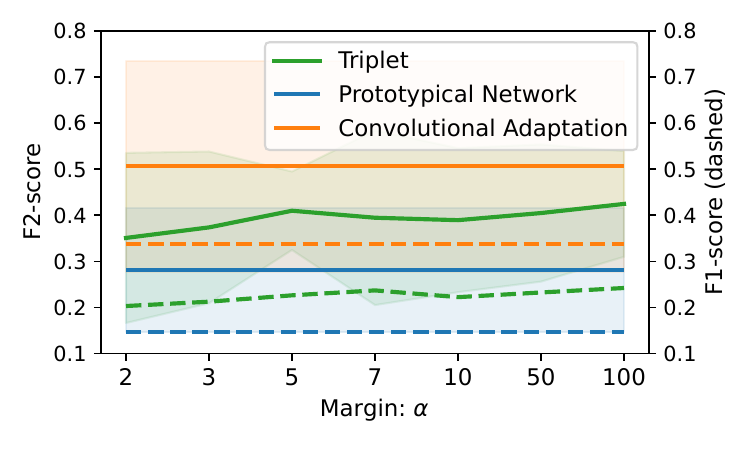}
        \subcaption{Triplet loss: $\alpha$.}
        \label{figure_hyper1}
    \end{minipage}
    \hfill
	\begin{minipage}[t]{0.493\linewidth}
        \centering
    	\includegraphics[trim=12 11 12 11, clip, width=1.0\linewidth]{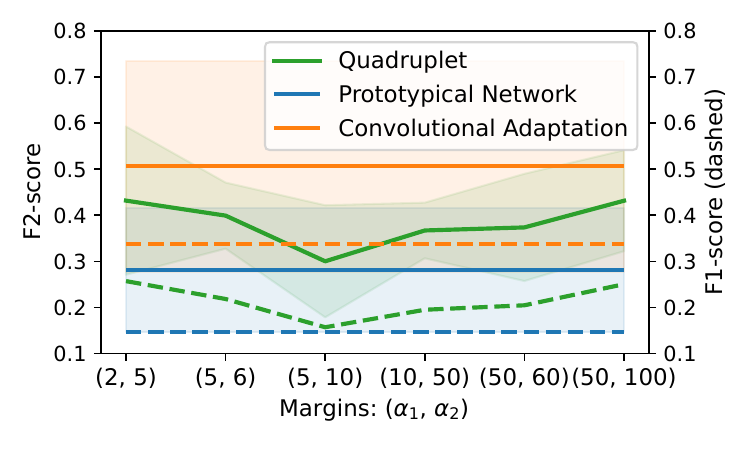}
        \subcaption{Quadruplet loss: $\alpha_1$ and $\alpha_2$.}
        \label{figure_hyper2}
    \end{minipage}
    \caption{Evaluation results (F2-score \& F1-score averaged over 10 trainings) for the hyperparameter searches of the margins.}
    \label{figure_hyper}
    \vspace{-0.2cm}
\end{figure}

\textbf{Pairwise Learning Results.} We first perform a hyperparameter search for the margins of the pairwise learning losses. For the triplet loss, we evaluate $\alpha \in \{2, 3, 5, 7, 10, 50, 100\}$ (see Figure~\ref{figure_hyper1}). For the quadruplet loss, we evaluate the pairs $(\alpha_1, \alpha_2)$ of both margins $(2, 5), (5, 6), (5, 10), (10, 50), (50, 60)$, and $(50, 100)$. Notably, the F2-score decreases for a low margin for the triplet loss and reaches its optimum with $\alpha=100$. For the quadruplet loss, the results are vice versa: We achieve the best results for $(2, 5)$ and $(50, 100)$. While we achieve an F2-score of 0.363 for the PN, the convolutional adaptation layer without post-training significantly drops with an 0.281 F2-score, but increases to 0.520 F2-score due to the availability of additional training data (averaged over $M=10$ trainings). Triplet loss further increases the results with 0.424 F2-score, respectively with an 0.431 F2-score for the quadruplet loss. The quadruplet loss leads to a higher generalizability with a more continuous representation. Please be advised that the F2-score serves as a stringent evaluation metric, particularly due to the unbalanced dataset (see Table~\ref{table_number_samples}). Despite the quadruplet loss yields an accuracy of 97.66\% and a recall of 0.929, the F2-score drops to 0.431 due to a notably low precision of 0.151. Although the quadruplet loss mitigates false positives, it results in a low false negative rate, i.e., high recall. We attain a high accuracy in classifying jammers across eight distinct classes. We select one specific permutation from Figure~\ref{figure_post_train_dense}, and evaluate the confusion matrix and uncertainties in Figure~\ref{figure_confusion_matrices} in more detail for the layer adaptation (left) and our quadruplet-based method (right). In general, the adaptation to the classes 7 to 9 is accurate due to non-similar samples (see Figure~\ref{figure_exemplary_samples}). We identify a higher aleatoric uncertainty (Figure~\ref{figure_confusion_matrices2}) and epistemic uncertainty (Figure~\ref{figure_confusion_matrices3}) for the class label pairs 0 and 3, as well as for the pairs 1, 2, 3, and 6, and the pairs 9 and 10. This leads to a higher confusion (Figure~\ref{figure_confusion_matrices1}) between the classes 1, 2, 3, 4, 5, and 6 as well as between 8, 9, and 10. With the quadruplet loss, we reduce the confusion of the classes 3, 6, and 8, at the cost of a lower accuracy for the negative class 0. Due to the more continuous representation, the aleatoric and epistemic uncertainty increases. The training time of the quadruplet loss marginally increases to 11.4 hours with 200 epochs.

\begin{figure}[!t]
    \centering
	\begin{minipage}[t]{0.493\linewidth}
        \centering
    	\includegraphics[trim=38 28 10 10, clip, width=1.0\linewidth]{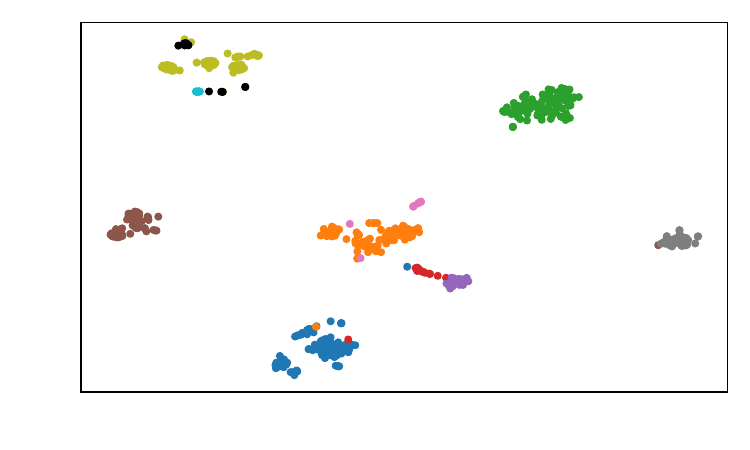}
        \subcaption{Prototypical network (PN).}
        \label{figure_embedding1}
    \end{minipage}
    \hfill
	\begin{minipage}[t]{0.493\linewidth}
        \centering
    	\includegraphics[trim=38 28 10 10, clip, width=1.0\linewidth]{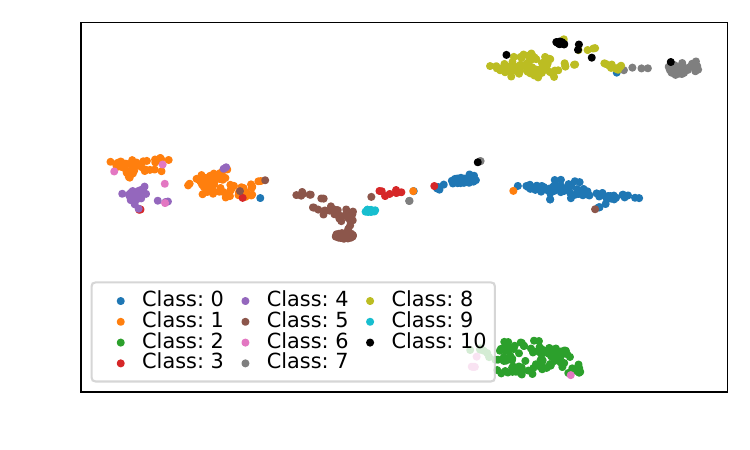}
        \subcaption{PN with quadruplet loss.}
        \label{figure_embedding2}
    \end{minipage}
    \caption{Embeddings of the 11 classes using t-SNE~\cite{maaten_hinton}.}
    \label{figure_embedding}
    \vspace{-0.2cm}
\end{figure}

\textbf{Embedding Analysis.} Figure~\ref{figure_embedding} illustrates the feature embedding of the last convolutional layer, which has an output size of 512, utilizing the t-distributed stochastic neighbor embedding (t-SNE) \cite{maaten_hinton} with perplexity of 30, an initial momentum of 0.5, and a final momentum of 0.8. We incorporate all 11 classes for the PN (refer to Figure~\ref{figure_embedding1}) and the PN trained with the quadruplet loss (refer to Figure~\ref{figure_embedding2}). We observe that for the PN, the classes are very distant in the lower dimension. However, the classes 3, 4, and 6, as well as 8 and 10 cannot be separated well. In the case of training with the quadruplet loss, there is a better transition between similar classes, i.e., 1, 3, 5, and 9. Consequently, the quadruplet loss leads to a more continuous representation.
\section{Conclusion}
\label{label_conclusion}

We introduced a FSL method designed for interference classification in GNSS data, to adapt an ML model to unseen jammer types with limited labels. The primary contribution is an uncertainty-based quadruplet loss, which enhances model generalization. We conducted extensive evaluations on our real-world dataset collected along a motorway. The quadruplet-based FSL method achieves an accuracy of 97.66\% and an F2-score of 0.431 that shows superior performance compared to the PN (0.363) and the triplet loss (0.424).

\section*{Acknowledgments}
This work has been carried out within the DARCII project, funding code 50NA2401, sponsored by the German Federal Ministry for Economic Affairs and Climate Action (BMWK) and supported by the German Aerospace Center (DLR), the Bundesnetzagentur (BNetzA), and the Federal Agency for Cartography and Geodesy (BKG).

\bibliography{references}
\bibliographystyle{IEEEtran}

\end{document}